\documentclass[longauth,letter]{aa}  
\usepackage{graphicx}
\usepackage{natbib}
\usepackage{color}
\usepackage{soul}

\newcommand{\CII}{[C\,{\sc ii}] }

\newcommand{\OI}{[O\,{\sc i}] }
\newcommand{\kms}{km~s$^{-1}$}

\usepackage{txfonts}
\begin{document}
\title{Gas morphology and energetics at the surface of PDRs:  new insights  with {\it Herschel}
observations of NGC 7023\thanks{{\it Herschel } is an ESA space observatory with 
science instruments 
provided by European-led Principal Investigator consortia with 
important participation from NASA.}}
\author{
C.~Joblin\inst{1,2} \and
P.~Pilleri\inst{1,2} \and
J.~Montillaud\inst{1,2} \and
A.~Fuente\inst{3} \and
M.~Gerin\inst{4} \and
O.~Bern\'e\inst{5} Ê\and
V.~Ossenkopf\inst{6,7} \and
J.~Le Bourlot\inst{8} \and
D~.Teyssier\inst{9} \and
J.~R.~Goicoechea\inst{10} \and
F.~Le Petit\inst{8} \and
M.~R\"ollig\inst{6} \and
M.~Akyilmaz\inst{6} \and
A.O.~Benz\inst{11} \and
F.~Boulanger\inst{12} \and
S.~Bruderer\inst{11} \and
C.~Dedes\inst{11} \and
K.~France\inst{13} \and
R.~G\"usten\inst{14} \and
A.~Harris\inst{15} \and
T.~Klein\inst{14} \and
C.~Kramer\inst{16} \and
S.D.~Lord\inst{17} \and
P.G~Martin\inst{13} \and
J.~Martin-Pintado\inst{10} \and
B.~Mookerjea\inst{18} \and
Y.~Okada\inst{6} \and
T.G.~Phillips\inst{19} \and
J.R.~Rizzo\inst{10} \and
R.~Simon\inst{6} \and
J.~Stutzki\inst{6} \and
F.~van der Tak\inst{7,20} \and
H.W.~Yorke\inst{21} \and
E.~Steinmetz\inst{22} \and
C.~Jarchow\inst{22} \and
P.~Hartogh\inst{22} \and
C.E.~Honingh\inst{6} \and
O.~Siebertz\inst{6} \and
E.~Caux\inst{1,2} \and
B.~Colin\inst{17}
 }

\institute{
Universit\'e de Toulouse, UPS, CESR, 9 avenue du colonel Roche, 31028 Toulouse cedex 4, France
\and
CNRS, UMR 5187, 31028 Toulouse, France
\and
Observatorio Astron\'omico Nacional (OAN), Apdo. 112, 28803 Alcal\'a de Henares (Madrid), Spain
\and
LERMA, Observatoire de Paris, 61 Av. de l'Observatoire, 75014 Paris, France
\and
Leiden Observatory, Universiteit Leiden, P.O. Box 9513, NL-2300 RA Leiden, The Netherlands
\and
I. Physikalisches Institut der Universit\"at
zu K\"oln, Z\"ulpicher Stra\ss{}e 77, 50937 K\"oln, Germany
\and
SRON Netherlands Institute for Space Research, P.O. Box 800, 9700 AV
Groningen, Netherlands
\and
Observatoire de Paris, LUTH and Universit\'e Denis Diderot, Place J. Janssen, 92190 Meudon, France
\and
European Space Astronomy Centre, Urb. Villafranca del Castillo, P.O. Box 50727, Madrid 28080, Spain
\and
Centro de Astrobiolog\'ia, CSIC-INTA, 28850, Madrid, Spain
\and
Institute for Astronomy, ETH Z\"urich, 8093 Z\"urich, Switzerland
\and
Institut d'Astrophysique Spatiale, Universit\'e Paris-Sud, B\^at. 121, 91405 Orsay Cedex, France
\and
Department of Astronomy and Astrophysics, University of Toronto, 60 St. George Street, Toronto, ON M5S 3H8, Canada
\and
Max-Planck-Institut f\"ur Radioastronomie, Auf dem H\"ugel 69, 53121, Bonn, Germany
\and
Astronomy Department, University of Maryland, College Park, MD 20742, USA
\and
Instituto de Radio Astronom\'ia Milim\'etrica (IRAM), Avenida Divina Pastora 7, Local 20, 18012 Granada, Spain
\and
IPAC/Caltech, MS 100-22, Pasadena, CA 91125, USA
\and
Tata Institute of Fundamental Research (TIFR), Homi Bhabha Road, Mumbai 400005, India
\and
California Institute of Technology, 320-47, Pasadena, CA Ê91125-4700, USA
\and
Kapteyn Astronomical Institute, University of Groningen, PO box 800, 9700 AV Groningen, Netherlands
\and
Jet Propulsion Laboratory, Caltech, Pasadena, CA 91109, USA
\and
MPI f\"ur Sonnensystemforschung, D 37191 Katlenburg-Lindau, Germany
}

\authorrunning {C. Joblin et al. } 
\titlerunning{Gas morphology and energetics at the surface of PDRs: NGC~7023 }

\abstract
{We investigate the physics and chemistry
of the gas and dust in dense photon-dominated regions (PDRs), along with
their dependence on the illuminating UV field. }
{Using {\it Herschel}-HIFI observations, we study the gas energetics in NGC~7023
in relation to the morphology of this nebula. NGC~7023 is the prototype of a
PDR illuminated by a B2V star and is one of the key targets of {\it Herschel}.}
{Our approach consists in determining the energetics of the region by
combining the information carried by the mid-IR spectrum
(extinction by classical grains, emission from very small dust particles) with that of the main gas coolant
lines. In this letter, we discuss more specifically the
intensity and line profile of the 158\,$\mu$m (1901\,GHz) \CII line measured by HIFI
and provide information on the emitting gas.
}
{We show that both the \CII emission and the mid-IR emission from polycyclic aromatic hydrocarbons (PAHs)
arise from the regions located in the transition zone between atomic and molecular gas.
Using the Meudon PDR code and a simple transfer model, we find good agreement between the
calculated and observed \CII intensities.}
{HIFI observations of NGC~7023 provide the opportunity to constrain
the energetics at the surface of PDRs. Future work will include analysis of the main coolant
line \OI and use of a new PDR model
that includes PAH-related species.Ê}


\maketitle

\section{Introduction}

One main goal of the Guarantee Time Key Programme "Warm and dense
interstellar medium" (WADI) of  the HIFI heterodyne spectrometer  \citep{degrauw10}
onboard {\it Herschel } \citep{pilbratt10} 
is to investigate the physics and chemistry
of the gas and dust in dense photon-dominated regions (PDRs), as well as
their dependence on the illuminating UV field. As part of this programme,
we observed a prototype PDR, NGC~7023.
The region is  illuminated by the B2Ve HD~200775
$[$RA(2000) = 21h01m36.9s ; Dec(2000) = +68$^\circ$09${'}$47.8${''}$],
 and has been shaped by the star formation process leading to the formation of a cavity. NGC~7023  has been widely studied
at many wavelengths. It has been shown that this region hosts structures at different gas densities:
$n_H\sim100$\,cm$^{-3}$ in the cavity, $\sim 10^4$\,cm$^{-3}$ in the PDRs
that are located north-west (NW), south (S) and east, and 
$10^5-10^6$\,cm$^{-3}$  in dense filaments and clumps that are observed in the mm (\citealt{fuente96, gerin98}
and references therein) and near-IR  \citep{lemaire96, martini97}.

NGC~7023 has been mapped by the instruments PACS and SPIRE of  {\it Herschel } to study the emission of large cold grains \citep{abergel10}.
We present here some observations of the gas at the surface of this nebula,
taking advantage of the very high spectral resolution of HIFI.
By combining these observations with previous mid-IR observations,
we study the geometry and energetics of the NW and S PDRs.

\begin{figure}
\centering
\includegraphics[ trim=1cm 0cm 2cm 1.5cm,clip, width=1 \columnwidth]{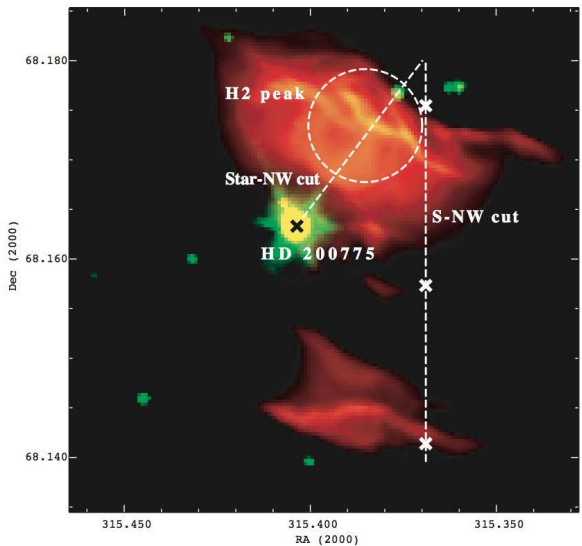}
\caption{The NW and S PDRs of NGC~7023 observed by {\it Spitzer}-IRAC at 8 $\mu$m (red) and 3.6 $\mu$m (green).
The white circle represents the HIFI beam at 535 GHz (41\arcsec) towards the H$_2$ peak. The dotted lines show the cuts that are studied in the \CII emission line at
158\,$\mu$m with a beam of 11\arcsec, whereas white crosses indicate the specific positions reported in Fig. \ref{fig:profiles}
 and Table \ref{tab_observation}. The star position is shown with a black cross. }
\label{fig:map}
 \end{figure}

\begin{figure}
\centering
\includegraphics[trim=0cm 0cm 0cm 1cm , height = 5.5cm , width=0.49 \columnwidth]{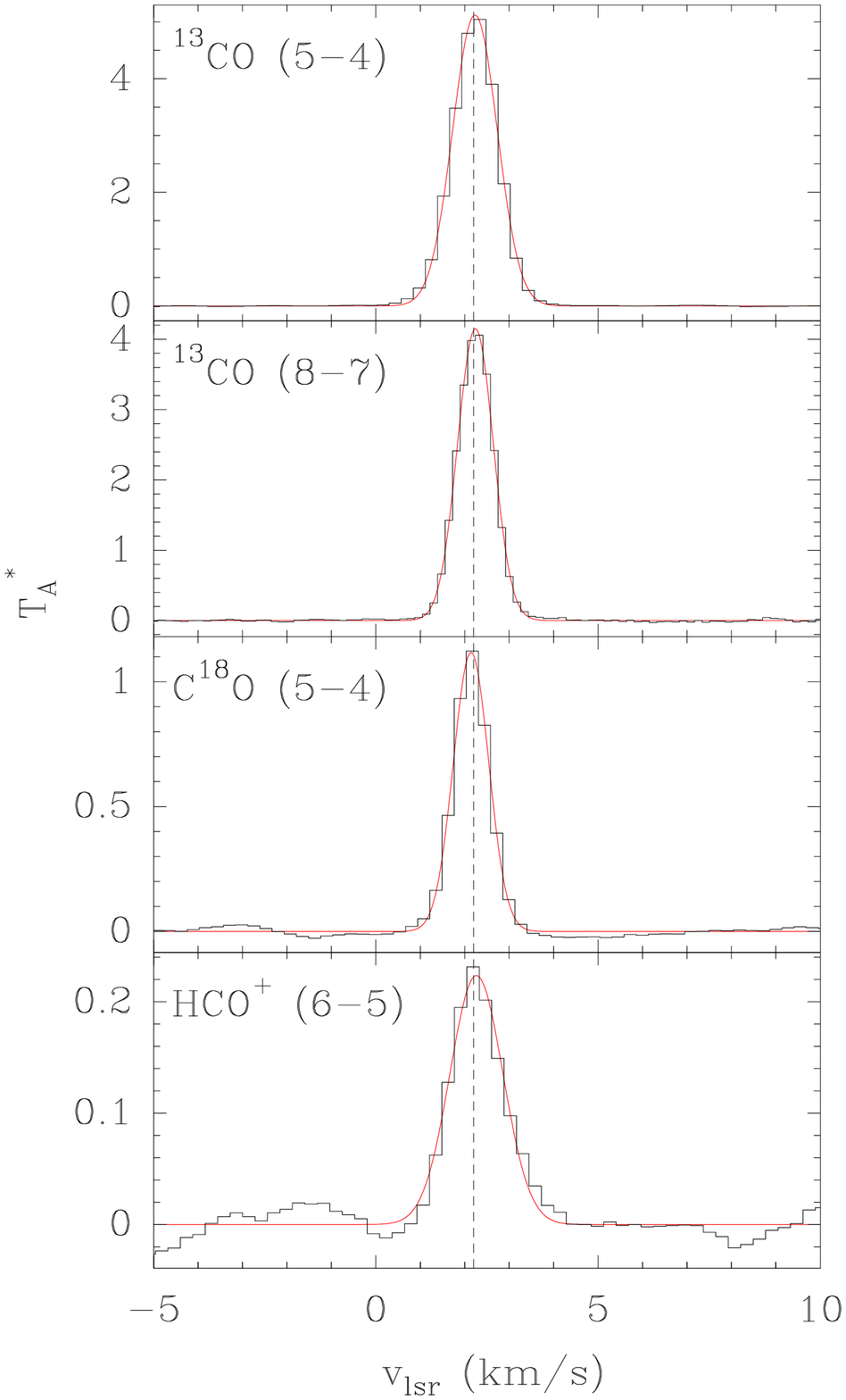}
\includegraphics[width=0.49 \columnwidth]{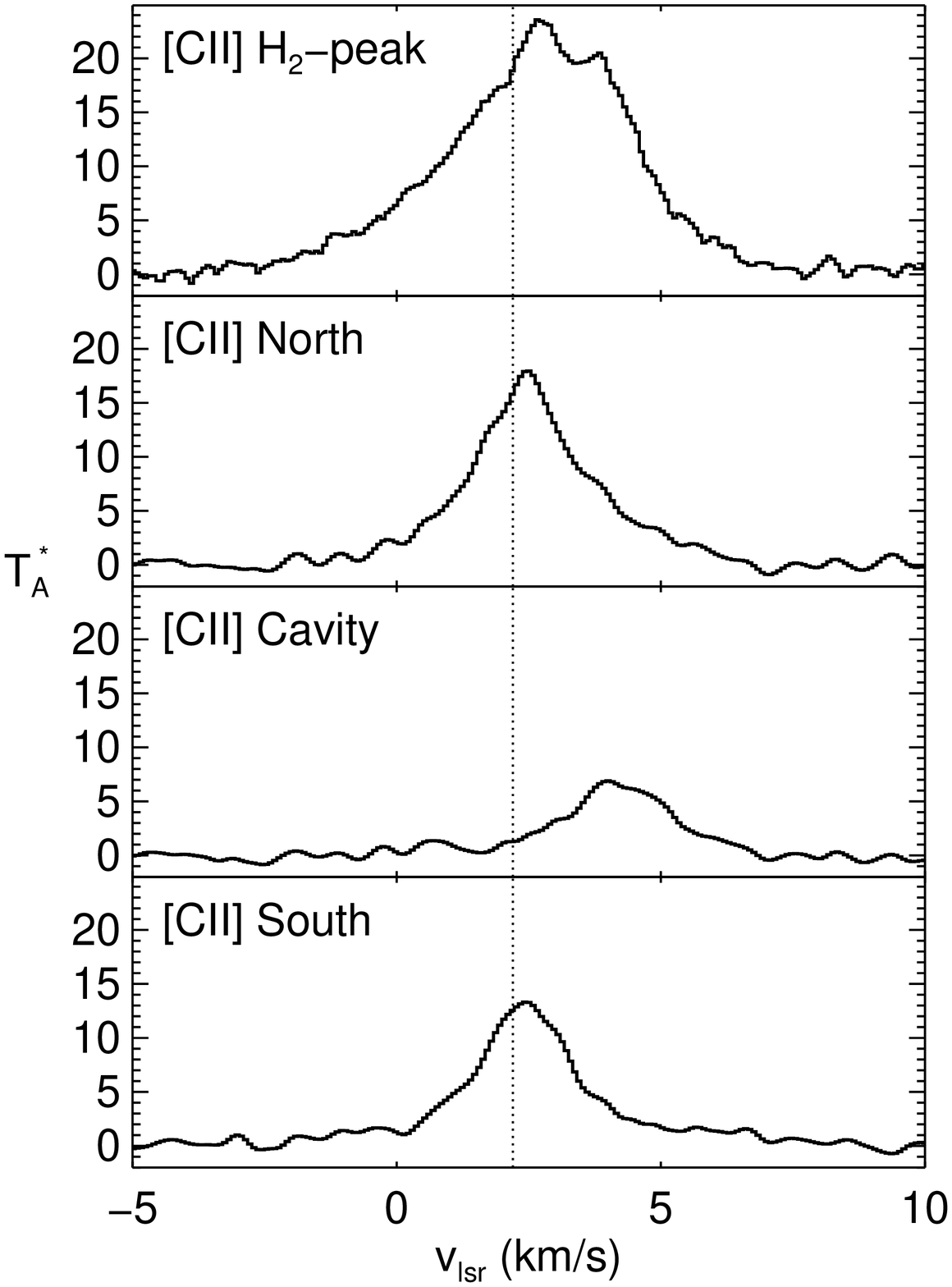}
\caption{{\it{Left:}} $^{13}$CO, C$^{18}$O, and HCO$^+$
high-J transitions observed with HIFI toward the H$_2$ peak. {\it Right:} Examples of
\CII emission profiles at the H$_2$ peak and at the three positions shown by crosses on Fig.\,\ref{fig:map}: 
north ($\Delta\alpha = -47\arcsec$, $\Delta\delta$ = +45\arcsec),  cavity  ($\Delta\alpha = -47\arcsec$, $\Delta\delta $=-20\arcsec ),
and south  ($\Delta\alpha = -47\arcsec$, $\Delta\delta$ = -80\arcsec ), with offsets relative to the star position. In both figures, the vertical dotted line indicates v$_{lsr}$ = 2.2 \kms.}
\label{fig:profiles}
\end{figure}

\begin{table*}
\caption{Summary of  the HIFI data}
\label{tab_observation}
\begin{center}
\begin{tabular}{lrlccccc}
\hline
Transition &     $\nu_{line}$ & Position &    Beam &     v$_{lsr}$ & FWHM $^\dagger$&     T$_A^*$ &    Area        \\
&     $[$GHz$ $] & &     size $[$\arcsec$]$ &    $[$\kms$]$ &     $[$\kms$]$ &     $[$K$]$ &    $[$ K\,km\,s$^{-1}]$    \\
\hline
HCO$^+$ (6$-$5) &     535.062 & H$_2$ peak &    41 &     $2.3 \pm 0.3$ & $1.4\pm0.3$ &     0.22 &    $0.33 \pm 0.07$    \\
$^{13}$CO (5$-$4) &      550.926 & H$_2$ peak &    41 &     $2.2 \pm 0.3$ & $1.2\pm0.3$ &      5.12 &    $6.5 \pm 1.5$    \\
$^{13}$CO (8$-$7) &      881.272 & H$_2$ peak &    24 &     $2.2 \pm 0.2$ & $0.9\pm0.2$ &     4.15 &    $4.0 \pm 0.8$    \\
C$^{18}$O (5$-$4) &     548.831 & H$_2$ peak &    41 &      $2.1 \pm 0.3$ & $0.9\pm0.3$ &     1.12 &    $1.1 \pm 0.4$     \\
\CII{} &     1900.537 & H$_2$ peak &    11 &     $2.7 \pm 0.1$ & $3.4 \pm 0.2$ &     23.6 &    $85.6 \pm 5.0$\\
\CII{} &     1900.537 & S-NW cut / North &    11 &     $2.5 \pm 0.1$ & $2.4\pm 0.2$ &     17.9 &    $45.8 \pm 3.8$    \\
\CII{} &     1900.537 & S-NW cut / Cavity &    11 &     $4.0 \pm 0.1$ & $2.2 \pm 0.2$ &    6.89 &    $16.2 \pm    1.4$\\
\CII{} &     1900.537 & S-NW cut / South &    11 &     $2.5 \pm 0.1$ & $2.4 \pm 0.2$ &     13.3 &    $34.0 \pm 2.8$\\
\hline
\multicolumn{8}{l}{$^\dagger$ For \CII, we report the FWHM of the Gaussian profile of equivalent area and peak intensity.}
\end{tabular}
\end{center}
\end{table*} 

\section{Observations and results}

\subsection{HIFI observations}

The HIFI observations presented here consist in (offset positions are relative to the star; see Fig. \ref{fig:map}):\\
- single pointing using frequency switch mode towards the NW PDR in bands 1a and 3b.
In band 1a, the frequency ranges covered by the Wide Band Spectrometer (WBS) were
[535-539]~GHz (LSB) and [547-551]~GHz (USB).
In band 3b, covered ranges were [879-883]~GHz (LSB) and~[891-895]~GHz (USB).
The observed position ($\Delta\alpha = -25\arcsec$, $\Delta\delta = +38\arcsec$,
called H$_2$~peak) corresponds to the peak intensity of the H$_2$
ro-vibrational emission associated to the near-IR filaments \citep{lemaire96}. \\
-ÊOn-the-fly (OTF) mapping of the  \CII 1901\,GHz emission line in band 7b. 
The line was covered by both the WBS and the high-resolution spectrometer (HRS) in USB.
Two cuts were performed: a cut from the star to the NW PDR (star-NW cut) and a south-north cut (hereafter S-NW cut) Êcovering the
NW PDR, the cavity and the S PDR ($\Delta\alpha = -47\arcsec$, $-85 < \Delta\delta < +60\arcsec$, see
Fig. \ref{fig:map}). ÊThe pixel size after regriding is 6.5$\arcsec$.

All these observations include an OFF reference position in the western lobe of
the cavity ($\Delta\alpha = -144\arcsec$, $\Delta\delta = -47\arcsec$ ).
Data was reduced with HIPE 3.0 \citep{ott10} on level-2  data
produced with the standard pipeline. For the pointed observations in bands 1a
and 3b, manual steps consisted in stitching sub-bands, baseline removal, and correction for main beam efficiency
($\eta_{mb}$ =  0.71). The \CII WBS spectra required defringing, which was performed with standard 
HIPE tasks, and the best data quality was produced with the subtraction of two sinusoidal 
fringes. 
To verify the biases introduced by the fringe removal, we compared the WBS and HRS 
profiles, which showed good agreement both in profile and absolute intensity except for the weakest
lines (T$_a^* \lesssim 4$\,K).

\subsection{ Gas kinematics}
\label{sec:kinematics}
Figure \ref{fig:profiles} (left panel) shows the $^{13}$CO 5$-$4, $^{13}$CO 8$-$7, C$^{18}$O 5$-$4, and HCO$^+$ 6$-$5 lines  observed by HIFI towards the ÊH$_2$ peak. All the lines have a central velocity of about  2.2\,\kms, comparable to previous ground-based
observations in several molecular lines \citep{fuente93}. 
Figure \ref{fig:profiles} (right panel) shows the \CII line profiles at the H$_2$ peak and at different positions along the S-NW 
cut. 
The line is much broader than molecular lines and its profile shows a complex multi-component structure.
 The observations towards the PDRs show that the emission peaks at Êintermediate velocities (v$_{lsr}\sim$1.8-2.8\,\kms), 
 which have already been observed towards the NW PDR
in several molecular tracers \citep[see, for example, ][]{fuente96}. There is also a contribution
from  higher velocity components (v$_{lsr}\sim$4\,\kms), which dominate the emission in the cavity.

\subsection{Emission from very small dust particles and C$^+$}
\label{AIB}

 \begin{figure}
\centering
\includegraphics[trim=1.5cm 12.5cm 4cm 2cm,clip,width=9 cm]{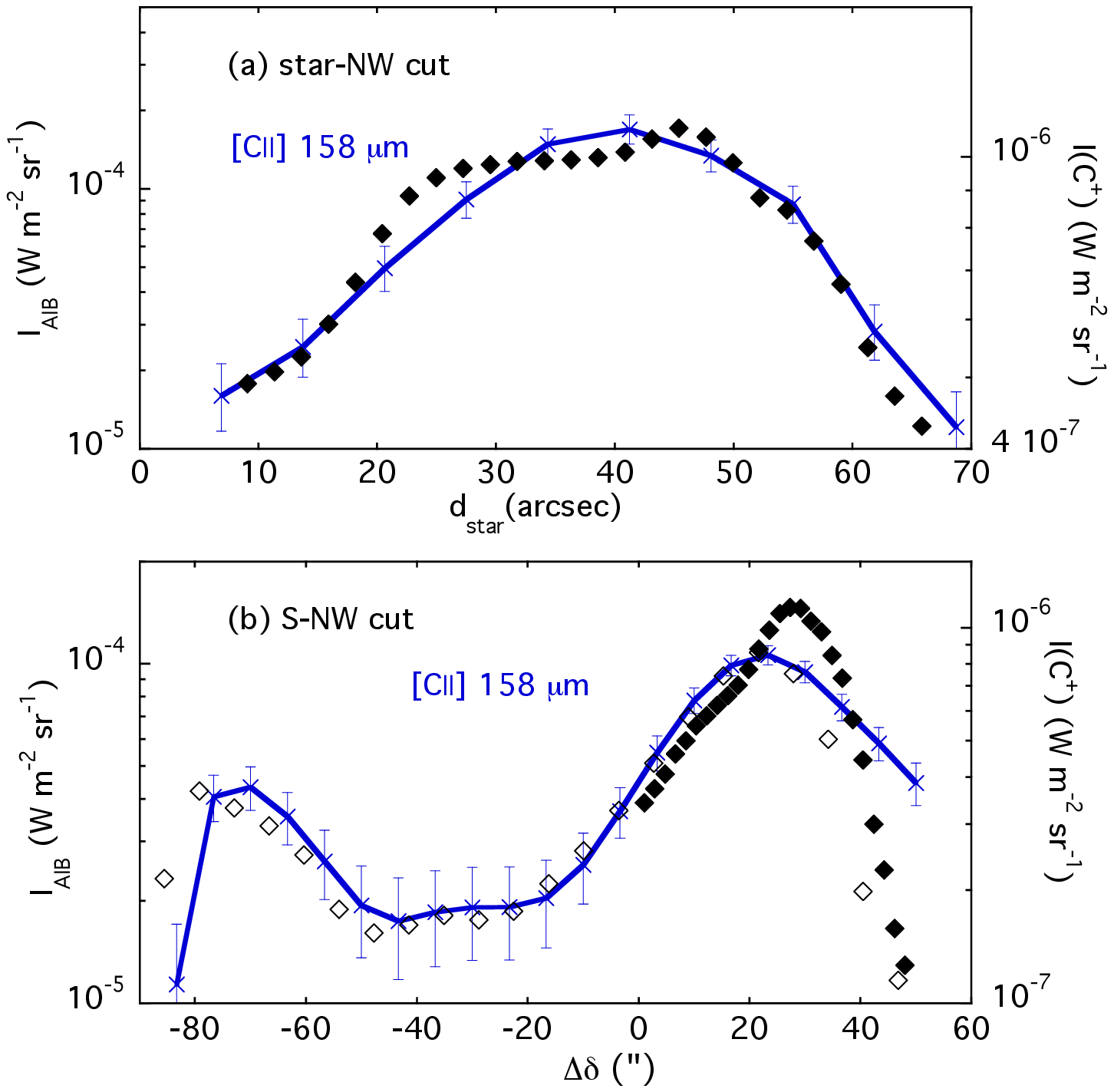}
\caption{Comparison between the  \CII 158\,$\mu$m line flux (solid line) measured with HIFI at a beam size of 11\arcsec\ and the aromatic IR band (AIB) flux (5.5-14\,$\mu$m) along the star-NW (a) and S-NW (b) cuts. The error
bars for \CII are computed at one-sigma level. The AIB flux is determined with a fit of the mid-IR spectra using the three PAH-related populations shown in Fig.\ref{fig:components};
filled diamonds are {\it Spitzer} data (1.8\arcsec pix$^{-1}$), and open diamonds  are ISOCAM data  (6\arcsec pix$^{-1}$).}
\label{fig:AIB_C+}
 \end{figure}
 
\begin{figure}
\centering
\includegraphics[trim=1.5cm 12.5cm 4cm 2cm,clip,width=9 cm]{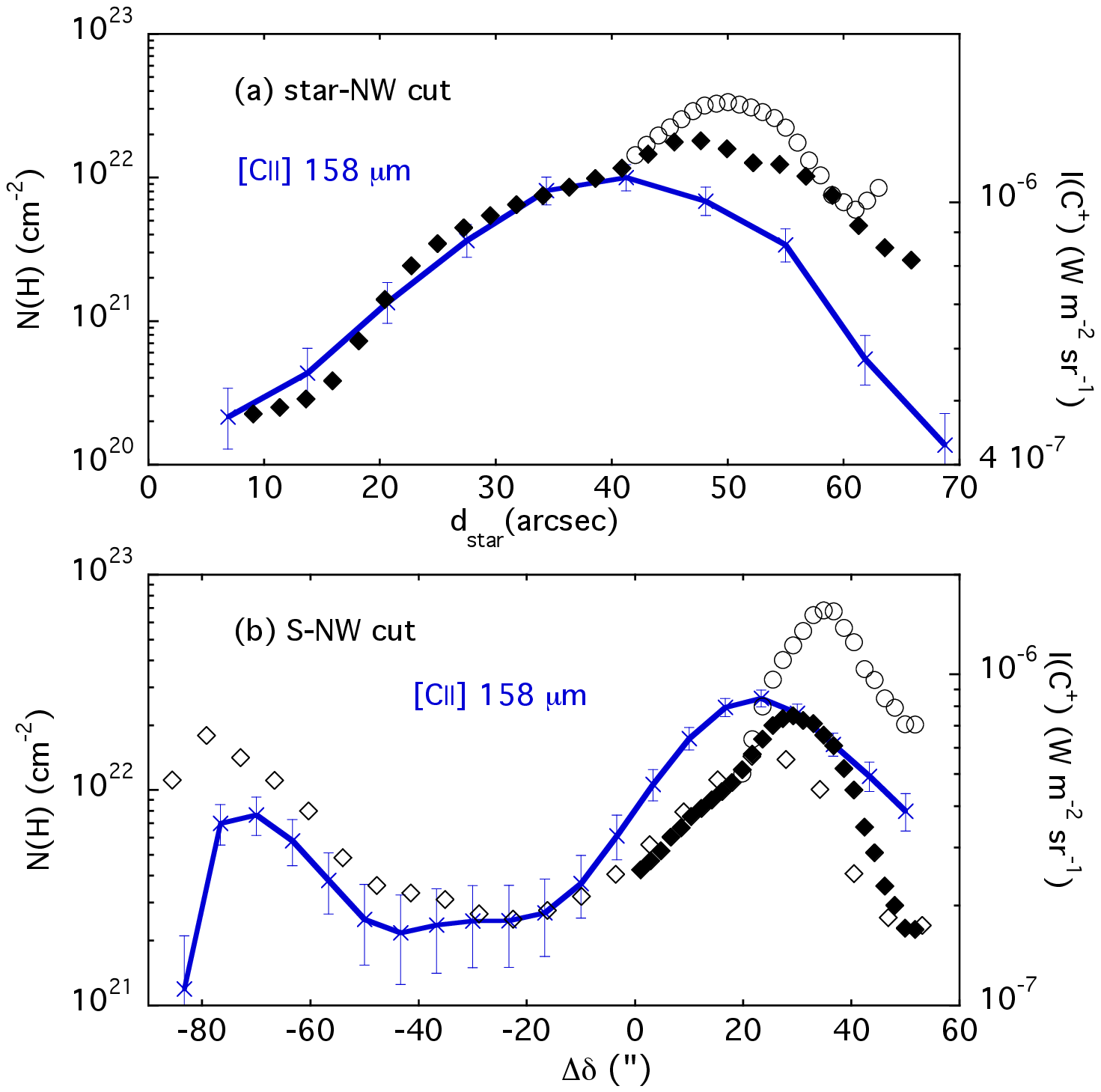}
\caption{Comparison between the  \CII 158\,$\mu$m line flux (solid line) measured with HIFI at a beam size of 11\arcsec\ and the
column density N(H) along the star-NW (a) and S-NW (b) cuts.  N(H) was derived from both the AIB flux (diamonds) and the
mid-IR dust extinction (open circles); filled diamonds and open circles are {\it Spitzer} data (1.8\arcsec pix$^{-1}$),
and open diamonds  are ISOCAM data  (6\arcsec pix$^{-1}$).}
\label{fig:NH_C+}
\end{figure}

The 158\,$\mu$m (1901\,GHz) \CII and 63\,$\mu$m \OI lines are the major coolants of the gas at the surface of PDRs
\citep{hollen89}.
 In these regions, photoelectric effect dominates the heating, while H$_2$ formation provides a minor contribution. Since the smallest dust particles, polycyclic aromatic hydrocarbons (PAHs) and very small grains (VSGs) contribute to a large fraction to this process \citep{bakes94, habart01}, and these particles emit in the mid-IR most of the energy they absorb in the UV, we expect that the mid-IR and \CII emissions arise in the same regions. 
 
We used mid-IR spectro-imagery data of the NW PDR of NGC~7023 that were obtained in the 5.5-14.5\,$\mu$m range with the Infrared Spectrograph onboard {\it Spitzer}
\citep{werner04}. For the S PDR, we used ISOCAM highly-processed data products \citep{boulanger96} from the ISO data archive. To analyse the mid-IR spectra, we followed the method explained in \citet{rapacioli05} and  \citet{berne07}, 
in which the mid-IR emission is decomposed into three aromatic IR band (AIB) spectra whose carriers are neutral PAHs (PAH$^0$), cationic PAHs (PAH$^+$), and evaporating VSGs.
The fitting procedure was recently improved by including the convolution of the composite spectrum by extinction  \citep{pilleri10}.
Figure\,\ref{fig:AIB_C+} displays the AIB flux, I$_{AIB}$, obtained by summing the fluxes of the PAH$^0$, PAH$^+$, and VSG components that were derived from the fit. It shows that the AIB intensity correlates well with the \CII  line intensity, strongly suggesting that both emissions arise from the same regions.
The fit of the mid-IR emission provides two independent tracers of the total gas column density N(H) along the line of sight as explained below.\\
(i)~ Owing to the excitation mechanism, the AIB intensity can be written as  I$_{AIB}\propto G_{0}\times N^C_{AIB}$
where N$^C_{AIB}$ is the column density of carbon in the AIB carriers and G$_0$ the UV flux in Habing units \citep{habing68}.
Assuming that N$^C_{AIB}/N(H)$ stays constant at the PDR surface, I$_{AIB}$ can therefore be used as a tracer of N(H) if G$_0$ is known \citep{pilleri10}.\\
(ii)~If the column density N(H) is high enough, the effect of extinction by silicates can be seen on the AIB spectrum. In the mid-IR fit, the extinction is derived from a simple correction term, assuming that the emitting and absorbing materials are fully mixed:
 $\frac{1-e^{-\tau_{\lambda} } }{\tau_{\lambda}} $ where $\tau_{\lambda} = N(H)\,C_{ext}
(\lambda)$  is the optical depth in the line of sight. The extinction cross-section per 
nucleon $C_{ext}(\lambda)$ is taken from \citet{weingartner01} for $R_V = 5.5$ and N(H) is a free parameter of the fit. \\
Method (ii)  is precise for column densities higher than
$N(H)\sim10^{22}$\,cm$^{-2}$. Method (i) can probe lower column densities but has two limitations.  The AIB emission needs to be corrected for the variation in the UV field G$_0$
to retrieve the value of N(H).  This was done assuming that G$_0$ scales as the inverse squared  distance to
the illuminating star HD200775 and a  value of G$_0$=2600 at 42$\arcsec$ from this star  \citep{pilleri10}. We used the projected distance as an estimate of the true distance.
This introduces an error that can be especially strong at  positions close to the star in the plane of the sky.
Figure 3 shows that the AIB emission stays almost constant at $d < 16$\arcsec, therefore we used this value as the minimum effective distance of the NW PDR to the star.
Method (i) also needs to be calibrated since the local emissivity of the AIB carriers is not known precisely. Our approach was therefore to derive a calibration factor using the values obtained by
method (ii) around position 42\arcsec on the  star-NW  cut. The same calibration factor was used for all positions along the two cuts. Figure\,\ref{fig:NH_C+} shows that the column densities
that were derived on the two cuts correlate quite well with the  \CII line intensity.

\section{Modelling C$^+$ emission}

\begin{table*}[t]
\caption{Summary of the PDR modelling of the \CII emission for 5 points along the HIFI S-NW cut ($\Delta\alpha = -47\arcsec$)}
   \begin{center}
   \begin{tabular}{r c c c c c c  c c c}
   \hline
    Pos. ($\Delta\delta$) & d$_{proj}$   &  UV field &   T  & PAH$^+$/PAH$^0$ &  Ionization &  $N(H)$    &  \CII local emissivity   & \multicolumn{2}{c}{\CII flux} \\
                          &            &     $\bf{(a)}$     & $\bf{(b)}$         &  $\bf{(c)}$  & parameter ($\gamma$) $\bf{(b)}$ & $\bf{(d)}$  & $\bf{(b)}$ &  HIFI & Model $\bf{(b)}$\\   
               (\arcsec)  &  (\arcsec) & (G$_0$)  &  (K) &                           & ($10^3$ G$_0$ K$^{1/2}$ cm$^{3}$)
                       &   ($10^{21} $ cm$^{-2}$)   &   (10$^{-21}$W m$^{-3}$)  &  \multicolumn{2}{c}{($10^{-7}$W m$^{-2}$ sr$^{-1}$)} \\
   \hline
   NW3 (16)    & 50  &  1873  &  337 / 333   &  1.9   &  11.0 / 29.5   & 10.5 &  4.4 / 1.4   & 8.0 & 11.0 / 8.0 \\
   NW2 (12)    & 48  &  1975  &  342 / 333   &  2.7   &  11.7 / 31.1   &  8.1 &  4.4 / 1.4  & 6.9 &  9.5 / 7.1 \\
   NW1 (-3)    & 47  &  2100  &  348 / 333   &  9.6   &  12.6 / 33.0   &  3.9 &  4.4 / 1.4  & 3.3 &  5.6 / 4.6 \\
   S1 (-63)    & 79  &   747  &  248 / 320   &  1.7   &  3.8  / 11.8   &  8.7 &  4.2 / 1.4  & 3.1 &  8.3 / 7.2 \\
   S2 (-73)    & 87  &   607  &  230 / 312   &  0.96  &  3.0  /  9.5   & 14.0 &  4.1 / 1.4  & 3.7 &  9.6 / 8.4 \\
   \hline   
 \multicolumn{10}{l}{$\bf{(a)}$ Calculated using a projected distance and G$_0$=2600 at 42\arcsec\ from the star.} \\
 \multicolumn{10}{l}{$\bf{(b)}$ From the PDR model using $n_H = 2\times10^4 /~7\times10^3\,$cm$^{-3}$, respectively.}\\
 \multicolumn{10}{l}{$\bf{(c)}$ Given as the ratio of the mid-IR intensities shown in Fig.\,\ref{fig:components}.}\\
   \multicolumn{10}{l}{$\bf{(d)}$ Derived from the analysis of the mid-IR emission spectra.}
   \label{model}
   \end{tabular}
   \end{center}
\end{table*}

The critical density for the  \CII 158\,$\mu$m line is n$_{crit}$=2500\,cm$^{-3}$ for collisions with H and therefore
the line emissivity depends mainly on temperature for n$>$n$_{crit}$. We selected a few positions on
the HIFI S-NW \CII cut, three points on the NW PDR and two on the South PDR (cf. Table\,\ref{model}).
The values of G$_0$ were determined as explained in Sect.\,\ref{AIB}, 
and we assumed a constant average density with two different values: $n_H = 2 \times 10^4$ cm$^{-3}$ that is characteristic
of the molecular cloud \citep{gerin98} and $n_H = 7 \times 10^3$ cm$^{-3}$ that was derived by \citet{rapacioli06} in their study of
PAH-related species. 

We used the 1D Meudon PDR code \citep{lepetit06} to compute the gas temperature T at the cloud surface for all the
selected positions (cf. Table\,\ref{model}). The values of T are used to calculate the C$^+$ level populations.
Line intensities are then derived by integrating along the line-of-sight (perpendicular to model results) and by
assuming uniform excitation conditions. 
The thickness of the observed regions leads to an optical depth $\tau \sim$1, which implies that transfer effects must be taken into account.
If we assume constant excitation conditions and gas properties along the line of sight, then $\tau$ and the line intensity can be computed. 

The agreement between calculated and observed flux values is very good when using $n_H = 7\times10^3$cm$^{-3}$.
In the NW PDR, the ratio is 1.0 for NW3 (16) and NW2 (12), and 1.4 for NW1 (-3).
For the S PDR, a value of 2.3 is derived for the two positions, suggesting that systematic effects are causing
the deviation between observed and calculated values of the \CII flux.
There are several parameters that are not precise in our model but looking at Table\,\ref{model}, it seems the  local \CII emissivity
is mainly affected by the local density and not by the value of G$_0$.  For N(H), we assumed the same regions emit in PAHs and \CII, in agreement with the profiles shown in Fig.\,\ref{fig:AIB_C+}.
There is also an error for N(H) due to our method (cf. Sect.\,\ref{AIB}), but this error is expected to be the same for both PDRs. 
Dividing N(H) by a factor of two leads to lower values of the ratio of the calculated over the observed [CII] flux: 0.7-0.8
for the NW PDR and 1.6-1.7 for the S PDR.

\begin{figure}
\centering
\includegraphics[trim=1.5cm 20cm 4cm 2.3cm,clip,width=9 cm]{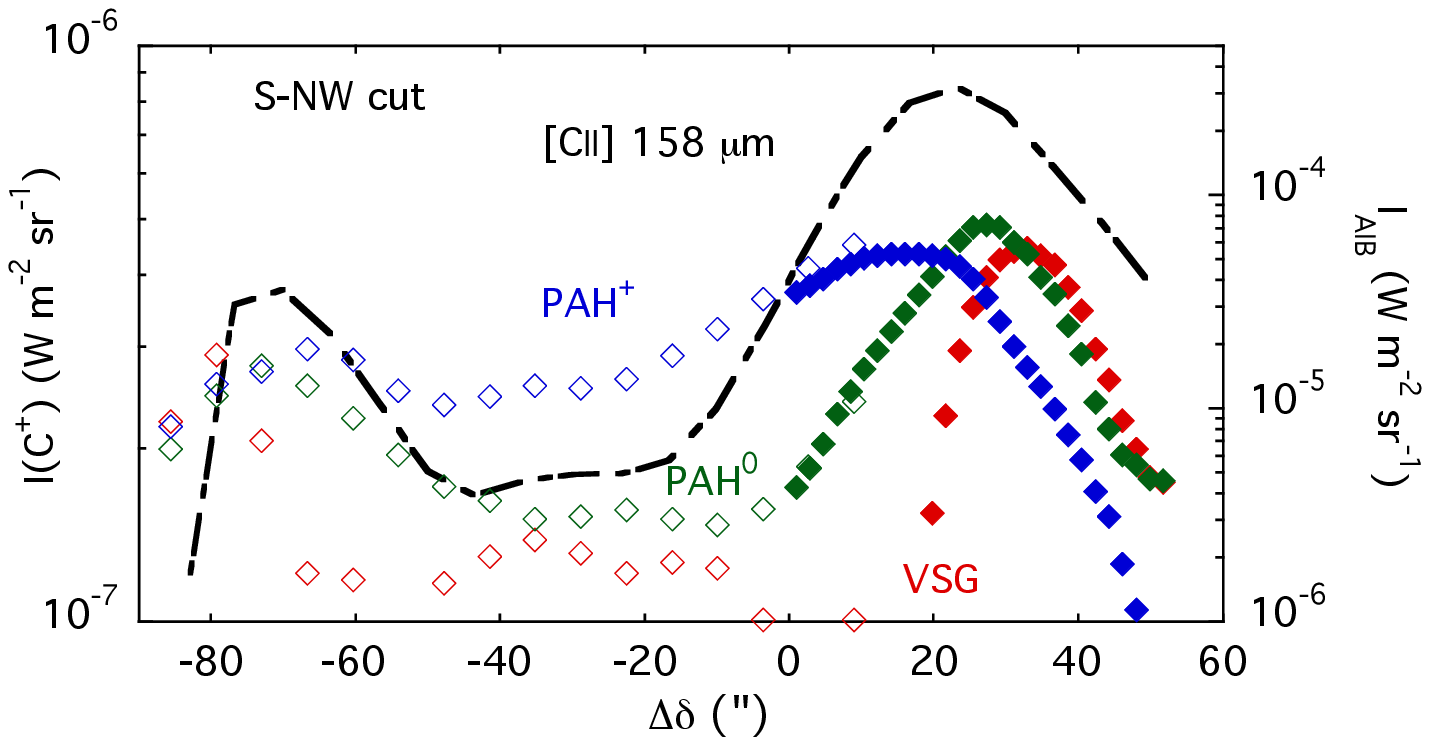}
\caption{ S-NW cut in \CII emission measured with HIFI at a beam size of 11\arcsec (dot-dashed line) and distribution of the emission from the different
very small dust populations, PAH$^+$ (blue), PAH$^0$ (green) and evaporating VSGs (red); filled diamonds are {\it Spitzer} data (1.8\arcsec pix$^{-1}$) and open diamonds are ISOCAM data  (6\arcsec pix$^{-1}$).}
\label{fig:components}
 \end{figure}

One step further in the model would consist in studying the effect of the grain charge on the photoelectric efficiency \citep{bakes94}. The relative abundances of PAH$^+$, PAH$^0$, and evaporating VSGs vary significantly over the nebula (Fig.\,\ref{fig:components}). Regions in the cavity appear mainly populated by PAH$^+$ (cf. NW1 (-3) in Table\,\ref{model}).
Since the ionization potential of PAH$^+$ is much higher than that of PAH$^0$ ($\sim$10\,eV compared to $\sim$6\,eV; \citealt{malloci07}),  PAH$^+$ should contribute less to the photoelectric heating than PAH$^0$, leading to a decrease in the heating rate, hence in the gas cooling.  In its current version, the PDR code uses classical grains with an MRN distribution \citep{mathis77} and absorption and scattering cross-sections from \citet{laor93}. We have used grains of sizes from 15\,\AA\ to 3000\,\AA\ with a dust-to-gas mass ratio of 1\%. As a result, the ionization parameter $\gamma$ that quantifies the grain charge (cf. Table\,\ref{model})  does not reflect well the variations of the PAH charge observed in Fig.\,\ref{fig:components}. An upgraded version, in which the PDR code is coupled to the code DUSTEM \citep{compiegne10}, is under development (Gonzalez et al., to be submitted) and will allow for including PAHs. NGC 7023 is clearly a template region that could be used for these studies.

\section{Conclusion}

By using HIFI and complementary mid-IR data, we have shown that the \CII cooling line and the AIB emission
arise from the same regions, in the transition zone between atomic and molecular gas. The prototype PDR NGC 7023
was found to be a good object for comparison with PDR models.
Further progress on the energetics of this region awaits for the coming \OI data from the PACS instrument and
a PDR model that treats the photophysics of PAHs consistently.



\begin{acknowledgements}
HIFI has been designed and built by a consortium of institutes and university departments from across Europe, Canada and the United States under the leadership of SRON Netherlands Institute for Space Research, Groningen, The Netherlands, and with major contributions from Germany, France, and the US. Consortium members are: Canada: CSA, U.Waterloo; France: CESR, LAB, LERMA, IRAM; Germany: KOSMA, MPIfR, MPS; Ireland, NUI Maynooth; Italy: ASI, IFSI-INAF, Osservatorio Astrofisico di Arcetri- INAF; Netherlands: SRON, TUD; Poland: CAMK, CBK; Spain: Observatorio Astron—mico Nacional (IGN), Centro de Astrobiolog'a (CSIC-INTA). Sweden: Chalmers University of Technology - MC2, RSS \& GARD; Onsala Space Observatory; Swedish National Space Board, Stockholm University - Stockholm Observatory; Switzerland: ETH Zurich, FHNW; USA: Caltech, JPL, NHSC. This work was supported by the German \emph{Deut\-sche For\-schungs\-ge\-mein\-schaft, DFG\/} project number Os~177/1--1. A portion of this research was performed at the Jet Propulsion Laboratory, California Institute of Technology, under contract with the National Aeronautics and Space administration.

\end{acknowledgements}

\end{document}